\documentclass{appolb}

\newcommand{\ima}{\hbox{Im}\,}
\newcommand{\rea}{\hbox{Re}\,}

\usepackage{graphicx}

\begin{document}
\title{Properties and composition of the $f_0(500)$ resonance
\thanks{Presented at ``Excited QCD''. Bjeliasnica Mountain, Sarajevo, 3-9 February 2013.}%
}
\author{J. R. Pel\'aez\footnote{speaker}, 
\vspace{-.4cm}
\address{Departamento de F\'{\i}sica Te\'orica II. Facultad de CC. F\'{\i}sicas. Universidad Complutense. 28040 Madrid. Spain}
\\ \vspace{.2cm}
J.~Nebreda
\vspace{-.4cm}
\address{Yukawa Institute for Theoretical Physics, Kyoto University, Kyoto 606-8502,
Japan}
\\ \vspace{.2cm}
G. R\'{\i}os and J. Ruiz de Elvira
\vspace{-.3cm}
\address{Helmholtz-Institut f\"ur Strahlen- und Kernphysik, Universit\"{a}t Bonn, D-53115 Bonn, Germany}
}
\maketitle
\begin{abstract}
\vspace{-.9cm}
In this talk we review our recent developments on the understanding of the
nature of the $f_0(500)$ resonance---or $\sigma$ meson---coming from the $N_c$
expansion, dispersion relations with Chiral Perturbation Theory, as well as finite energy sum rules and semi-local duality. 
\vspace{-.4cm}
\end{abstract}
  
\vspace{-.2cm}
\section{Introduction}
The $f_0(500)$ isoscalar-scalar resonance--also known as $\sigma$ meson--, or the correlated two-pion exchange with those quantum numbers, plays a prominent role in nucleon-nucleon attraction as well as in the spontaneous chiral symmetry breaking of QCD. Its properties and nature are therefore very relevant for Nuclear and Particle Physics, and it even has implications for Cosmological and Anthropic considerations. However, even though a scalar-isoscalar field 
 was first proposed \cite{Johnson:1955zz} more than 65 years ago to explain nucleon-nucleon attraction, its description in terms of quarks and gluons is still the subject of an intense debate.

Concerning the properties, namely its mass and width, it seems that the issue has been finally solved (see \cite{Pelaez:2013jp} for a brief review) using elaborated dispersive analyses of existing data, and this is why in the last edition of the Particle Data Tables (PDT)\cite{PDG}, the former $f_0(600)$ meson, previously quoted with a huge mass uncertainty from 400 to 1200 MeV, has changed name to $f_0(500)$ and is now quoted with a mass between  400 and 550 MeV. A similar reduction has taken place in the PDT for the width. Nevertheless, the very PDT suggest in their ``Note on light scalars below 2 GeV'' that 
one could ``take the more radical point of view and just
average the most advanced dispersive analyses'' \cite{CGL,Caprini:2005zr,GarciaMartin:2011jx,Moussallam:2011zg}, and obtain an estimated pole position
at $M-i\Gamma/2\simeq\sqrt{s_\sigma}=(446\pm6)-(276\pm5)\,$ MeV.

With the mass and width issue solved, we need to understand its composition in terms of quarks and gluons. Actually, the inverted mass hierarchy of the lightest scalar nonet, of which the $\sigma$ is the lightest member, hints at a possible tetraquark nature, as suggested long ago in \cite{Jaffe:1976ig}. However, thesigma nature is still a matter of intense debate today, due to the poor quality of the data available, the use of some strong model dependent analyses, and the additional complication of mixing between different configurations, which is expected to be very relevant in the meson sector. Nevertheless, in view of the supporting evidence, there is a growing consensus that the tetraquark---or possibly molecular---component might be dominant although it could be mixed with components of a different nature. The results of our group, that we review next, support this interpretation.

\vspace{-.4cm}
\section{The $f_0(500)$ predominant non-$\bar qq$ $ N_c$ behavior}

The $1/N_c$ expansion of QCD \cite{Witten:1980sp}, $N_c$ being the number of colors, can be applied at all energies (not only at high energies as the usual expansion in the coupling) and, for our purposes, provides a prediction for the leading order behavior of $\bar qq$ mesons, whose mass and width scale as $O(1)$ and $O(1/N_c)$, respectively. The behavior of other configurations is also known \cite{Witten:1980sp,Weinberg:2013cfa}.

\vspace{-.4cm} 
\subsection{Model Independent approach}

If the pole $s_R=m_R^2-im_R \Gamma_R$ of an elastic resonance (like the $\sigma$)
behaves as a $\bar qq$ 
then the scattering phase shift where it appears satisfies \cite{Nieves:2009kh}:
 \begin{eqnarray*}
 \delta(m_R^2)=\frac{\pi}{2}-\frac{\rea t^{-1}}{\sigma}
\Big\vert_{m_R^2}+O(N_c^{-3}), \quad
\delta'(m_R^2)=-\frac{(\rea t^{-1})'}{\sigma}
\Big\vert_{m_R^2}+O(N_c^{-1}), \nonumber
 \end{eqnarray*}
from which we have shown \cite{Nebreda:2011cp} that the adimensional observables 
\begin{eqnarray*}
\frac{\frac{\pi}{2}- \rea t^{-1}/\sigma}{\delta}\Big\vert_{m_R^2}
\equiv\Delta_1=1+\frac{a}{N_c^3}..., \qquad
-\frac{[\rea t^{-1}]'}{\delta'\sigma}\Big\vert_{m_R^2}\equiv\Delta_2=1+\frac{b}{N_c^2}....
\label{abdef}
\end{eqnarray*}
are equal to one up to corrections suppressed by more than just one power of $1/N_c$. Since $a,b$ are naturally expected to be of order one {\it or less} (cancellations with higher order terms can substantially decrease their effective value, but not increase it), and are coefficients of very suppressed corrections, they are
very sensitive to deviations from a $\bar qq$ behavior.

Now, we can use the scattering phase shifts obtained in \cite{GarciaMartin:2011cn}
within a model independent dispersive analysis of data and the $\sigma$ and $\rho(770)$ pole positions corresponding to that analysis, which are found in \cite{GarciaMartin:2011jx}. Altogether, these yield, for the $\rho(770)$ (widely accepted as an ordinary $\bar qq$ resonance): $a_\rho=-0.06\pm0.01$ and $b=0.37 ^{+0.04}_{-0.05}$, in good agreement with the expectations. In contrast, for the $\sigma$, we find: $a_\sigma=-252^{+119}_{-156}$ and $b_\sigma=77^{+28}_{-24}$. Two or more orders of magnitude larger than expected for a $\bar qq$. {\it This is a strong and model independent support for a predominant non-$\bar qq$ component for the $f_0(500)$}, since it only makes use of the QCD leading behavior of $\bar qq$ states and a dispersive data analysis.

A glueball component \cite{Nebreda:2011cp}, whose width scales as $1/N_c^2$, is more disfavored since the resulting $a,b$ are one order of magnitude larger. 

\vspace{-.4cm}
\subsection{Unitarized Chiral Perturbation Theory}

A different approach consists on using a partial wave dispersion relation for the inverse amplitude in order to describe the $\pi\pi$ scattering data. 
The elastic Inverse Amplitude Method (IAM)\cite{IAM} uses Chiral Perturbation Theory (ChPT)\cite{chpt} to evaluate the subtraction 
constants and the left cut of the dispersion relation. The elastic right cut is exact in the elastic approximation, since the
elastic unitarity condition $\ima t=\sigma |t|^2$,
fixes $\ima t^{-1}=-\sigma$. Note that the IAM is 
derived only from exact elastic 
unitarity, analyticity in the form of a dispersion
relation and ChPT, which is only used at low energies, and reproduces meson-meson
scattering data up to energies $\sim 1\,$GeV. It can be 
analytically continued into the second Riemann sheet to find the poles
associated to the
$\rho(770)$ and $f_0(500)$ resonances, which are generated from the unitarization process. 
The dependence 
on the QCD number of colors
is implemented \cite{Pelaez:2003dy,Pelaez:2006nj} through the
model independent leading $1/N_c$ scaling of the ChPT low energy constants (LECs) 
\cite{chpt,Pelaez:2006nj}.

\begin{figure}[htb]
\centerline{%
\includegraphics[width=5cm]{rhoPoleSU3cc.eps}
\includegraphics[width=8cm]{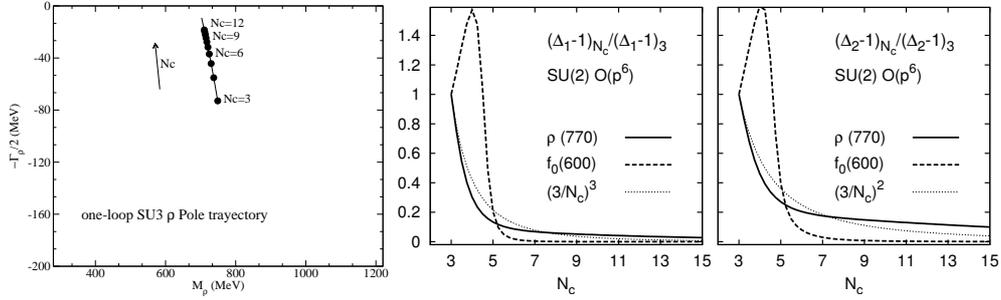}}
\caption{(Left) $N_c$ behavior of the $\rho(770)$ pole. (Center and Right) The $\Delta_1$ and $\Delta_2$ observables both for the $\rho(770)$ and $f_0(500)$,
versus the expected behavior for a $\bar qq$.}
\label{Fig:F1H}
\end{figure}

\begin{figure}[htb]
\centerline{%
\includegraphics[width=11cm]{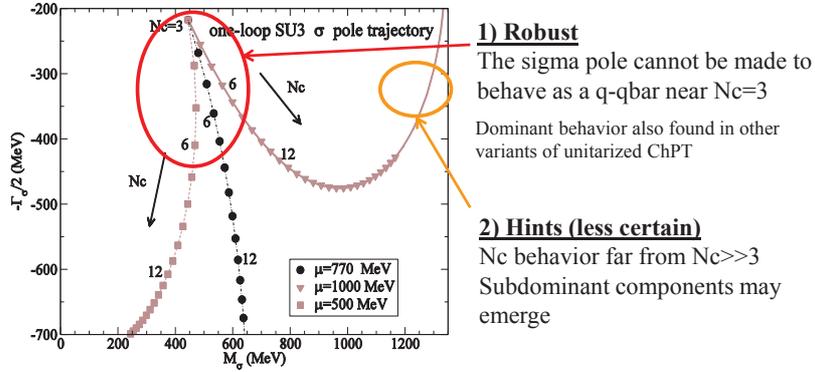}}
\caption{$N_c$ behavior of the $f_0(500)$ pole, including uncertainties.
Not far from $N_c=3$ the width increases dramatically with $N_c$, at odds
with a $\bar qq$ dominant component (compare the $\rho(770)$ in Fig.1). At two loops the preferred solution comes close to the $\mu=1000\,$MeV case.
The figure is from \cite{FESR}}
\label{Fig:F2H}
\end{figure}

Hence, by varying $N_c$ in the LECs, we obtain the $\rho(770)$ and $f_0(500)$ behavior. This can be done within ChPT to one-loop, $O(p^4)$ \cite{Pelaez:2003dy}, or two-loops, $O(p^6)$ \cite{Pelaez:2006nj}. Thus, we show in the left panel of Fig.\ref{Fig:F1H} the $\rho(770)$ pole movement in the complex plane. As expected for a $\bar qq$ its mass barely moves, whereas the width decreases with $1/N_c$. In the center and right panels we compare the behavior of the $\Delta_1$ and $\Delta_2$ observables defined above, versus their expected $1/N_c^3$ or $1/N_c$ behavior for ordinary mesons. Note the $\rho$ is very consistent with the expected behavior, but the $f_0(500)$ is completely at odds with it. Actually, from Fig.\ref{Fig:F2H}, which includes the uncertainty in the one-loop calculation, parameterized by the renormalization scale $\mu$ of the LECs \cite{Pelaez:2003dy,FESR}, we note that not far from $N_c=3$,  the width always grows as the pole moves deep in the lower complex plane as $N_c$ increases, in contrast to the shrinking width observed for the $\rho(770)$ in Fig.\ref{Fig:F1H}. This result, obtained not far from $N_c=3$ is very robust within uncertainties and has been confirmed by several authors \cite{otros}. Nevertheless, within the uncertainties, it is possible for the width to keep growing or decrease again. The latter could be interpreted as the effect of mixing with another $\bar qq$ component, which is subdominant in the $N_c=3$
physical world, but becomes dominant at larger $N_c$. Note however, that this subdominant $\bar qq$ component only arises {\it above 1 GeV} \cite{Pelaez:2006nj}, unless one spoils the data description or the $\rho(770)$ $\bar qq$ behavior.
This scenario could be expected within the most common interpretation, which suggests that the lightest scalar nonet is of a non-$\bar qq$ nature, whereas the first ordinary $\bar qq$ nonet appears around 1 to 1.5 GeV \cite{nonet}.
Furthermore, this ``subdominant $\bar qq$ component around 1-1.5 GeV'' scenario is favored by the two-loop analysis \cite{Pelaez:2006nj}. Moreover, our picture is qualitatively consistent
with recent lattice calculations \cite{Bali:2013kia} finding that
the lightest scalar is about a factor 1.5 heavier than the $\rho$ in the $N_c\rightarrow\infty$ limit.

\vspace{-.2cm}

\vspace{-.5cm}

\vspace{-.5cm}
\subsection{Semi local duality and the $N_c$ dependence of the $\rho(770)$ and $f_0(500)$}

In the real world ($N_c$ = 3) and  low energies, the scattering amplitude is well represented by a sum of resonances
(with a background), but as the energy increases, the resonances (having more phase space for decay)
become wider and increasingly overlap. This overlap generates a smooth Regge 
behavior described by a small number of crossed channel Regge exchanges. 
Thus, s-channel resonances
are related ``on the average'' to Regge exchanges in the t-channel, a feature known as ``semi-local duality''.

Now, data teach us that in the repulsive isospin I=2 
$\pi\pi$ scattering channel there are no resonances. Hence, 
semi-local
duality means that the I=2 t-channel amplitude should be small compared with other t-channels. But  the I=2 t-channel can be recast in terms of s-channel amplitudes as: 
\begin{equation}
\mathrm{Im}\,A^{t2}(s,t)=\frac{1}{3}\mathrm{Im}\,A^{s0}(s,t)-\frac{1}{2}\mathrm{Im}\,A^{s1}(s,t)+\frac{1}{6}\mathrm{Im}\,A^{s2}(s,t),
\end{equation}
$A^{s2}$ being small. Hence, to have a small $A^{t2}$ requires a strong cancellation between 
$A^{s0}$ and $A^{s1}$. However, these channels are saturated at low energies by 
the  $f_0$(500) and $\rho$(770) resonances, respectively. This ``on the average cancellation'' is properly defined via Finite Energy Sum
Rules:   
\begin{equation}\label{FESR}
F(t)^{21}_n= \frac{\int_{\nu_{th}}^{\nu_{\mathrm{max}}}{d\nu\;\mathrm{Im}\, A^{t2}(s,t)/\nu^n}}{\int_{\nu_{th}}^{\nu_{\mathrm{max}}}{d\nu\;\mathrm{Im}\,A^{t1}(s,t)/\nu^n}},\;\;\;\nu=(s-u)/2.
\end{equation}
Semi local duality implies $|F(t)^{21}_n| \ll 1$, which we have checked to be well satisfied for $n>1$ \cite{FESR}.
We expect the $I=2$ channel to remain repulsive and no resonances to appear even for $N_c\neq 3$. As a consequence {\it all models where the $\rho(770)$ and the $f_0(500)$ resonances behave differently}, are in potential conflict with semi-local duality.

\begin{figure}[htb]
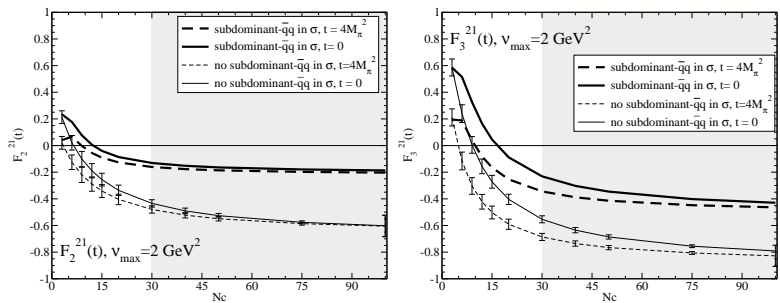

\centerline{%
\includegraphics[width=5cm]{R21n2-f2.eps}
\includegraphics[width=5cm]{R21n3-f2.eps}}
\caption{Semi-local duality is well satisfied,$|F(t)^{21}_n| \ll 1$, for all $N_c$ when a subdominant $\bar qq$ component around 1 to 1,5 GeV is present in the $f_0(500)$ \cite{FESR}.}
\label{Fig:F3H}
\end{figure}

We have recently shown \cite{FESR} that this conflict actually occurs in those scenarios where the $f_0(500)$ disappears deeply in the complex plane, \emph{but this conflict is not present when there is a subdominant $\bar qq$ component in the sigma in the 1 to 1.5 GeV region}. The results are summarized in Fig.\ref{Fig:F3H}.

\vspace{-.5cm}
\section{Acknowledgments}

\vspace{-.1cm}
J.R.P. thanks the organicers for creating such a lively and estimulating Workshop.
Work supported by the Spanish grant FPA2011-27853-C02-02.

\end{document}